\documentclass[useAMS,usenatbib]{mn2e}
\usepackage{graphicx}

\def\jj{\mbox{$J$}}

\def\ks{\mbox{$K_s$}}

\def\jk{\mbox{$(J-K_s)$}}

\def\aV{\mbox{$A_{V}$}}
\def\dV{\mbox{$A_V^{DR}$}}
\def\mMJ{\mbox{$(m-M)_J$}}
\def\eJK{\mbox{$E(J-K_s)$}}
\def\ms{\mbox{$\rm M_\odot$}}
\def\ds{\mbox{$d_\odot$}}

\def\ta{\mbox{$t_{age}$}}
\def\Mcl{\mbox{$M_{clu}$}}
\def\x2{\mbox{$R_{RMS}$}}
\def\fB{\mbox{$f_{bin}$}}

\title[Fundamental parameters of young star clusters]{Deriving reliable fundamental
parameters of PMS-rich star clusters affected by differential reddening}

\author[C. Bonatto, E. Bica and E.F. Lima]{C. Bonatto$^1$, E. Bica$^1$ and
E.F. Lima$^1$\\
$^1$ Departamento de Astronomia, Universidade Federal do Rio Grande 
do Sul, Av. Bento Gon\c{c}alves 9500\\
Porto Alegre 91501-970, RS, Brazil}

\begin{document}

\pagerange{\pageref{firstpage}--\pageref{lastpage}}

\maketitle

\label{firstpage}

\begin{abstract}
We present an approach that improves the search for reliable astrophysical parameters 
(e.g. age, mass, and distance) of differentially-reddened, pre-main sequence-rich star 
clusters. It involves simulating conditions related to the early-cluster phases, in 
particular the differential and foreground reddenings, and internal age spread. Given 
the loose constraints imposed by these factors, the derivation of parameters based only 
on photometry may be uncertain, especially for the poorly-populated clusters. We consider 
a wide range of cluster {\em (i)} mass and {\em (ii)} age, and different values of 
{\em (iii)} distance modulus, {\em (iv)} differential and {\em (v)} foreground reddenings. 
Photometric errors and their relation with magnitude are also taken into account. We
also investigate how the presence of unresolved binaries affect the derived parameters.
For each set of {\em (i)} - {\em (v)} we build the corresponding model Hess diagram, and 
compute the root mean squared residual with respect to the observed Hess diagram. The 
parameters that produce the minimum residuals between model and observed Hess diagrams 
are searched by exploring the full parameter space of {\em (i)} - {\em (v)} by means 
of {\em brute force}, which may be time consuming but efficient.  Control tests show that 
an adequate convergence is achieved allowing for solutions with residuals 10\% higher 
than the absolute minimum. Compared to a colour-magnitude diagram containing only 
single stars, the presence of 100\% of unresolved binaries has little effect on cluster 
age, foreground and differential reddenings; significant differences show up in the cluster 
mass and distance from the Sun. Our approach shows to be successful in minimising the 
subjectiveness when deriving fundamental parameters of young star clusters.
\end{abstract}


\section{Introduction}
\label{Intro}

Because of important dynamically-induced structural changes, the first few $10^7$\,yr 
represent the most critical period in a star cluster's life, especially for the 
low-mass embedded clusters (ECs). At this stage, cluster dissolution is essentially 
related to the impulsive parental gas removal by supernovae and massive star winds. 
Following the rapid change in the gravitational potential - and the reduced escape 
velocity - an important fraction of the stars, even all stars, escape to the field. 
This process has been shown capable of dissolving most of the very young star clusters 
on a time-scale of $10-40$\,Myr (e.g. \citealt{tutu78}; \citealt{GoBa06}). We point
out that, because of gas and dust ejection, young clusters with most of the stars still 
in the PMS are not necessarily ECs. Examples might be Bochum\,1 (\citealt{Bochum1}), 
Pismis\,5 (\citealt{Pi5}), and NGC\,4755 (\citealt{N4755}).

Current estimates (e.g. \citealt{LL2003}; \citealt{SFR}) suggest that less than $\sim5\%$ 
of the Galactic ECs dynamically evolve into gravitationally bound open clusters (OCs). Thus, 
such a massive early dissolution of ECs may be an important source of Galactic field stars 
(e.g. \citealt{Massey95}). 

It is in this context that a robust determination of fundamental parameters (e.g. 
age, distance, mass, reddening, etc) of young star clusters is important. However,
several factors severely challenge this task, most of which are related to the
conspicuous presence of pre-main sequence (PMS) stars in Colour-Magnitude Diagrams 
(CMDs) of young, low-mass star clusters. Evolving towards the under-populated MS, 
the PMS stars are usually shrouded within a non-uniform dust distribution, which may 
lead to a high degree of differential reddening. Examples of clusters characterised 
by such CMDs are NGC\,6611, NGC\,4755, NGC\,2244, Bochum\,1, Pismis\,5, NGC\,1931, 
vdB\,80, Cr\,197 and vdB\,92 (\citealt{N2244}; \citealt{vdB92} and references therein). 
This effect is particularly critical for the very young clusters that are still in the 
(gas and dust) embedded phase. As an additional complicating factor, star formation 
within a cluster is not characterised by a single event. Instead, stars in young 
clusters are observed to form over a significant time spread (e.g. \citealt{Stauffer97}, 
and references therein), usually comparable to the cluster age. In summary, CMDs of 
clusters younger than $\ta\sim30-40$\,Myr are expected to contain stars with a range 
of ages ($\le\ta$) and affected by varying degrees of differential reddening. And, 
since most of the stellar mass of a young cluster is stored in the PMS, the combined 
effect of the age spread and differential reddening complicates the straightforward 
derivation of cluster fundamental parameters. This is especially true when photometry 
is the only available information on a cluster. 

Given the relevance of the above issues, it's natural that previous approaches with 
different sophistication degrees have been developed, exploring similar lines as in the 
present paper. For instance, \citet{NJ06} present a powerful and formally elegant 
maximum-likelihood method to derive distances and ages of young clusters through 
comprehensive Hess diagram simulations. Although allowing for the presence of binaries
in an isochronal population, their method does not include age spread or differential 
reddening. Also, it appears to apply more consistently to clusters with CMDs satisfactorily
described by a single isochrone, i.e., those without a significant PMS age spread (usually 
older than $\sim30$\,Myr). Later, \citet{daRio2010} improves on the \citet{NJ06} method by 
including differential reddening, age spreads and PMS stars in the simulations. However,
distance and reddening are not free parameters in \citet{daRio2010}; instead, they are
adopted from previously estimated values. \citet{Hill08} presents another attempt to modelling 
CMDs of young star clusters by means of varying star formation histories. Based on confusion
between signal and noise in CMDs, they conclude that there is only marginal evidence for 
moderate age spreads in recent star forming regions and young open clusters. More recently,
\citet{StHo2011} apply a Monte Carlo method to the age determination of embedded clusters 
through near-infrared (UKIDSS) photometry. They deredden the photometry of a real cluster
and compare it with models built from theoretical isochrones.

In this work we simulate some relevant conditions that usually apply to the early cluster 
phases to approach the problem of obtaining reliable fundamental parameters of PMS-rich 
clusters affected by varying degrees of differential reddening. We adopt a simplistic 
approach, keeping the number of assumptions - and free parameters - to a minimum. In 
short, we start by building a distribution of artificial stars (i.e. with mass and 
absolute luminosity) corresponding to clusters of a range of masses and ages. Next, we 
apply a set of values of foreground reddening, distance modulus and differential 
reddening to the model stars, and build the corresponding Hess diagram. Photometric
uncertainties and their smearing effect on CMDs are explicitly taken into account. At each step 
we compare the artificial and observed Hess diagrams, searching for the set of values 
that produce the best match. Binaries - in varying fractions - are also included
in the simulated Hess diagrams. Formally, our {\em brute-force} approach is not as 
elegant as some of the previous methods (see above). However, it has the advantage of 
fully exploring the parameter space in the search for the best set of values, although 
at the cost of heavy computer time when working with fine grids (Sect.~\ref{Test}).

\begin{figure}
\resizebox{\hsize}{!}{\includegraphics{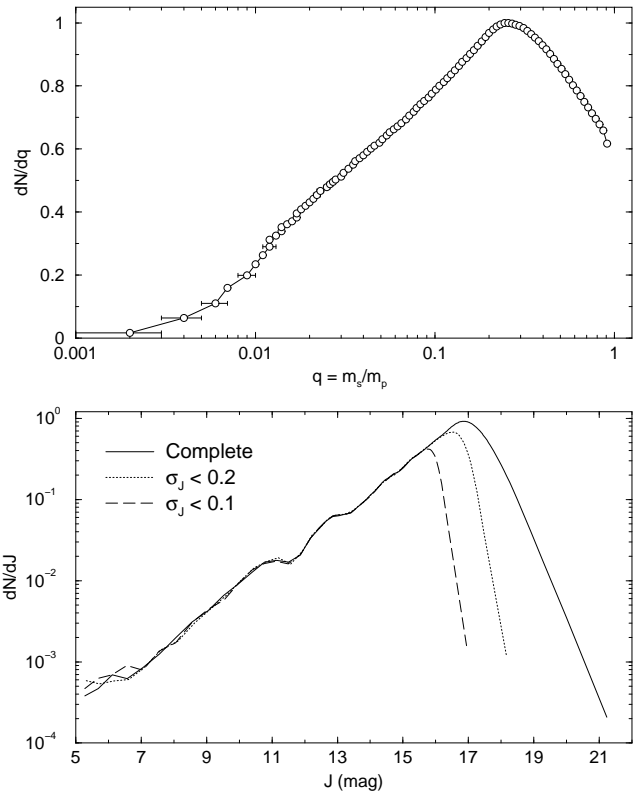}}
\caption[]{Top: Binary mass-ratio distribution that result from pairing stars with similar 
ages (and any mass) on a cluster undergoing continuous star formation for 20\,Myr. 
$m_s$ and $m_p$ are the secondary and primary-star mass, respectively. The distribution's
peak at $q\approx0.25$ has been normalized to unity. Bottom: \jj-band photometric completeness 
as a function of different error tolerances (see Sect.~\ref{Test}
for details).}
\label{figA}
\end{figure}

\begin{figure}
\resizebox{\hsize}{!}{\includegraphics{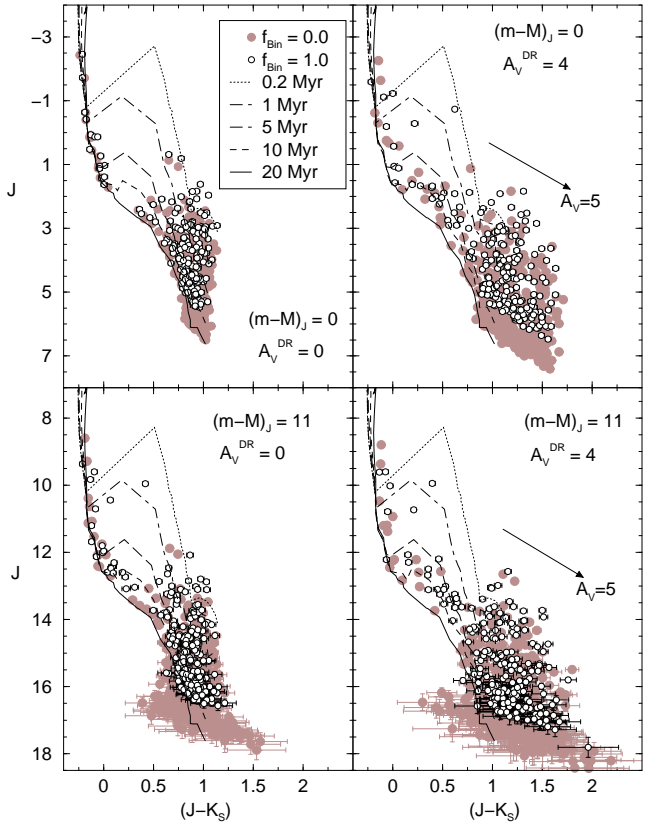}}
\caption[]{Template CMD of a 20\,Myr and 250\,\ms\ cluster with stars submitted to 
$\dV=0$ (left panels) and $\dV=4$ (right), displaced at $\mMJ=0$ (top) and $\mMJ=11$ 
(bottom). The arrow shows the reddening vector for $\aV=5$. The foreground reddening 
is zero in all cases. For comparison, evolutionary sequences corresponding to binary 
fractions of 0\% (filled symbols) and 100\% (empty) are considered. }
\label{fig1}
\end{figure}

The present paper is organised as follows. In Sect.~\ref{YCS} we discuss the relevant 
effects that affect the CMDs and describe the approach. In Sect.~\ref{ConExp} we apply 
the approach to some control cases and discuss the results in terms of the residual
statistics. In Sect.~\ref{TCADF} we do the same with two previously studied young 
clusters. Concluding remarks are given in Sect.~\ref{Conclu}.

\section{Young cluster simulations}
\label{YCS}

For consistency with previous work by our group on young clusters (e.g. \citealt{vdB92}), 
we consider here photometric properties (e.g. the relationship between errors with apparent 
magnitude, quality control, CMDs) that usually apply to 2MASS\footnote{The Two Micron All 
Sky Survey, All Sky data release (\citealt{2mass06})}. The all-sky coverage and uniformity 
of 2MASS, associated with a moderate near-infrared photometric depth, provide an adequate 
environment for probing properties even of deeply embedded clusters (e.g. \citealt{Sh252} 
and references therein). For similar reasons, we work with $\jj\times\jk$ CMDs (this colour 
is the least affected by photometric errors and the best discriminant for PMS stars
- e.g. \citealt{Teu34}) and the corresponding Hess diagrams. Thus, in what follows we refer 
to the apparent distance modulus as \mMJ, foreground reddening as \eJK, and differential 
reddening as \dV. Reddening transformations are based on the absorption relations 
$A_J/A_V=0.276$ and $A_{K_S}/A_V=0.118$, with $R_V=3.1$ (\citealt{Cardelli89}; \citealt{DSB2002}). 

Photometric uncertainties - which are assumed to be Normally distributed -  are taken into 
account when building the Hess diagrams. Formally, if the magnitude (or colour) of a given 
star is given by $\bar x\pm\sigma_x$, the probability of finding it at a specific value $x$ 
is given by $P(x)=\frac{1}{\sqrt{2\pi}\sigma_x}\,e^{{-\frac{1}{2}}\left(\frac{x-\bar x}
{\sigma_x}\right)^2}$. Thus, for each star we compute the fraction of the magnitude and 
colour that occurs in a given bin of a Hess diagram, which corresponds to the difference 
of the error functions computed at the bin borders. By definition, summing the colour and
magnitude density over all Hess bins results in the number of input stars. As a compromise
between CMD resolution and computational time, the Hess diagrams used in this work consist 
of magnitude and colour bins of size $\Delta\jj=0.2$ and $\Delta\jk=0.02$, respectively. 

Binary systems are expected to survive the early evolutionary phase of low-mass
clusters, with the unresolved pairs being somewhat brighter than the single stars and
producing some broadening of the CMD stellar sequences (e.g. \citealt{NJ06}). In addition,
binaries also produce changes on the initial mass function of young, massive star clusters
(e.g. \citealt{Weidner09}). Thus, we also include them in our simulations by means of
the parameter \fB, which measures the fraction of unresolved binaries in a CMD. According
to this definition, a CMD having $N_{CMD}$ detections, but characterised by the binary
fraction \fB, would have a number of individual stars expressed as $N_* = (1+\fB)N_{CMD}$.
For consistency with our assumption of continuous star formation (Sect.~\ref{Test}), 
binaries are formed by pairing stars with the closest ages, regardless of the individual
masses. This gives rise to a secondary to primary star mass-ratio ($q=m_s/m_p$) that 
smoothly increases from very-low values up to $q\approx0.25$, and decreases for higher 
values of $q$ (Fig.~\ref{figA}).

\subsection{A first look at the problem}
\label{ModPrem}

Most of the difficulties associated with obtaining reliable fundamental parameters for 
young clusters are summarised in Fig.~\ref{fig1}, in which we present a template CMD 
corresponding to a 250\,\ms\ cluster with an age of 20\,Myr, and no foreground 
reddening. For the stellar mass/luminosity relation we use the solar-metallicity
isochrone sets of Padova (\citealt{Girardi2002}) isochrones\footnote{Computed  for the 
2MASS filters at {\em http://stev.oapd.inaf.it/cgi-bin/cmd}.} and \citet{Siess2000}. 
Both isochrone sets have been merged for each age considered, since Padova isochrones 
should be used only for the MS (or more evolved sequences), while those of Siess apply to 
the PMS. The merging point occurs at the MS, at $6.5\,\ms$ for the isochrones
younger than 8\,Myr, $5.5\,\ms$ for 10\,Myr, $4.5\,\ms$ for 20\,Myr, and $3.5\,\ms$ for 
30\,Myr. The models are built with stars with mass $\ge0.1\,\ms$, the lowest mass 
considered in the PMS isochrones. 

First we consider zero distance modulus and no differential reddening. At this point the 
template stars are relatively bright and simply distribute among the isochrones according 
to the age and mass, with very small photometric uncertainties. When the template stars 
are displaced by e.g. $\mMJ=11$ (still with no differential reddening), a significant 
scatter shows up, increasing for fainter stars. Finally, when a moderate value of 
differential reddening ($\dV=4$) is added to the template, the scatter tends to mask 
any relationship between stars and isochrones, even for $\mMJ=0$. Clearly, this effect 
increases with distance modulus. At each step we also show the stellar sequences 
corresponding to the maximum possible binary fraction, $\fB=1.0$. While the mild 
brightening implied by the binaries with respect to the single stars (Sect.~\ref{TCADF}) 
is clearly seen in all cases, the broadening ends up drowned both by the intrinsic PMS 
age spread and photometric errors. Thus, the binary-related broadening should be more 
noticeable in the MS of massive clusters of any age.

Note that, as a photometric quality control, we have kept only the stars with \jj\ and 
\ks\ errors lower than 0.5. Such a loose constraint is used here only for illustrative 
purposes. In what follows we work only stars with errors lower than 0.2. The error
restriction is used to somehow emulate the photometric completeness function of the
observations. We illustrate this effect on the luminosity function of an artificial
cluster containing an arbitrarily large number of stars, for statistical purposes
(Fig.~\ref{figA}). After following the steps described below for assigning magnitudes
and uncertainties, we applied the restriction of keeping only stars with $\sigma_J\le
0.1$ and 0.2. Compared to the {\em complete} luminosity function, which presents a
turnover at $\jj\approx16.9$, the restricted functions have turnovers that smoothly
shift towards brighter magnitudes.

\subsection{Description of the approach}
\label{Test}

Briefly put, we start by building the Hess diagram corresponding to the $\jj\times\jk$ 
CMD of a young cluster. Then we search - by means of {\em brute force} - for the set of 
values of cluster stellar mass (\Mcl), age (\ta), differential reddening (\dV), 
foreground reddening (\eJK), and apparent distance modulus (\mMJ), that produces the 
best match between the observed and simulated Hess diagrams. Note that, to minimise
the number of free parameters, we assume a uniform (or flat) distribution for the 
differential reddening. Alternative shapes might be tested, such as a normal distribution
around a mean value. However, this would require an additional parameter (the standard 
deviation), and CMDs of young clusters might lack constraints to find the {\em best values} 
for a large number of free parameters. Also, we assume that for a cluster of age \ta\ and 
mass \Mcl, stars (of mass $m$) form continuously between $0\le t\le\ta$. In this context, 
the cluster age also characterises the star-formation time spread, since we assume the 
cluster age to coincide with the beginning of the star formation. As a caveat we
note that assuming a flat age distribution may be somewhat unrealistic, especially for
clusters older than $\sim10$\,Myr, since this would imply a very slow and steady star 
formation rate. However, except for the artificial clusters used as control tests below, 
in this paper we apply our approach to clusters younger than this threshold.

The approach takes on the following steps: {\em (i)} Start with an artificial cluster 
of mass \Mcl, age \ta, apparent distance modulus \mMJ, and foreground reddening \eJK. 
{\em (ii)} Select the stellar masses by randomly taking values from \citet{Kroupa2001} 
mass function, until the individual mass sum yields \Mcl. {\em (iii)} Randomly assign 
each star an age $\le\ta$. {\em (iv)} As another simplifying assumption, consider that 
each star can be randomly absorbed by any value of (differential) reddening in the range 
$0\le\aV\le\dV$. {\em (v)} Apply shifts in colour and magnitude according to the values 
of \eJK\ and \mMJ. {\em (vi)} Assign each artificial star a photometric uncertainty based 
on the average 2MASS errors and magnitude relationship. {\em (vii)} For more realistic 
representativeness, add some photometric noise to the stars. This step is done to 
minimise the probability of stars with the same mass having exactly the same {\em observed} 
magnitude, colour, and uncertainty in the CMD. Consider a star (of mass $m$ and age \ta) 
with an intrinsic (i.e., measured from the corresponding isochrone) magnitude $mag_i$ and 
assigned uncertainty $\sigma_{mag}$. The noise-added magnitude $mag$ is then randomly 
computed from a Normal distribution with a mean $mag_i$ and standard deviation $\sigma_{mag}$.
{\em (viii)} Apply the same detection limit to the model CMD as for the observations, 
so that model and data share a similar photometric completeness function; in practice, this 
means that stars with photometric errors higher than 0.2 are discarded. {\em (ix)} Build 
the corresponding $\jj\times\jk$ Hess diagram, compare it with the observed one and compute 
the root mean squared residual \x2\ (see below) for this set of values. {\em (x)} Repeat 
steps {\em (i)} - {\em (x)} for a range of values of \Mcl, \ta, \dV, \eJK, and \mMJ. 
Finally, analyse the topology of the hyperspace defined by $\x2=\x2(\Mcl, \ta, \dV, 
\eJK, \mMJ)$ and search for solutions around the minimum values of \x2. Some technical 
details are described below.

\citet{Kroupa2001} mass function is defined as $dN/dm\propto m^{-(1+\chi)}$, with the 
slopes $\chi=0.3$ for $0.08\leq m(\ms)\leq0.5$ and $\chi=1.3$ for $m(\ms)>0.5$. The
relation of errors with magnitude for 2MASS is well represented by $\sigma_J = 0.0214 + 
2.48\times10^{-8}\exp{(J/1.071)}$ and $\sigma_{K_s}=0.0193+9.59\times10^{-9}\exp{(K_s/1.067)}$. 
As for the photometric noise for a star with magnitude $mag\pm\sigma_{mag}$, a new magnitude is 
randomly taken from the Normal distribution characterised by the mean $mag$ and standard 
deviation $\sigma_{mag}$.

With respect to the root mean squared residual between observed ($H_{obs}$) and 
simulated ($H_{sim}$) Hess diagrams composed of $n_c$ and $n_m$ colour and magnitude 
bins, we define 

$$\x2=\sqrt{\sum_{i,j=1}^{n_c,n_m}\frac{\left[H_{obs}(i,j)-H_{sim}(i,j)\right]^2}
{n_c\times n_m}}.$$

To minimise the critical stochasticity associated with low-mass clusters, in step {\em (i)} 
we build $N_{sim}$ clusters of mass \Mcl\ and age \ta. To improve the statistical significance, 
the final simulations, in most cases, are run with $N_{sim}=250$. In this sense, the artificial 
Hess diagram (step {\em (ix)}) corresponds to the average density over the $N_{sim}$ clusters.

\begin{figure}
\resizebox{\hsize}{!}{\includegraphics{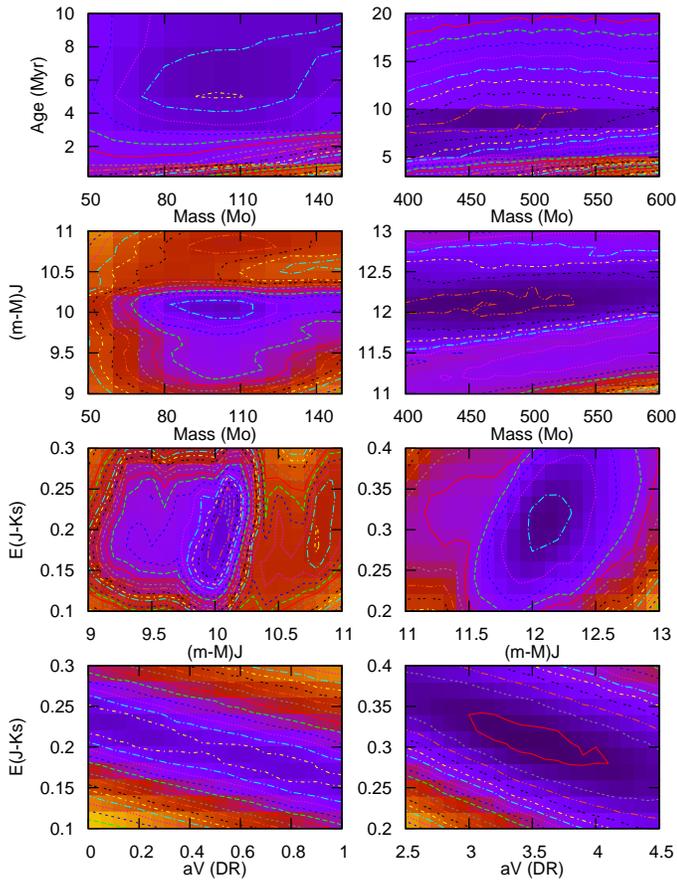}}
\caption[]{\x2\ contour maps for selected parameters of MODEL\#2 (left)
and MODEL\#3 (right). Darker colours indicate lower values of \x2. The absolute 
minima are $\x2=0.034$ (MODEL\#2) and $\x2=0.028$ (MODEL\#3).}
\label{fig2}
\end{figure}

\begin{figure}
\resizebox{\hsize}{!}{\includegraphics{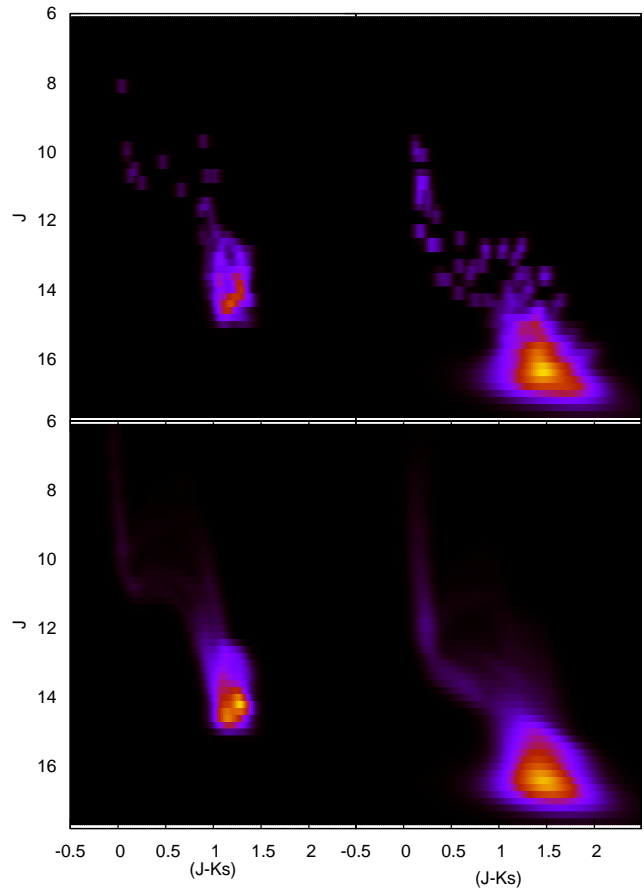}}
\caption[]{Observed (top panels) and simulated (bottom) Hess diagrams for MODEL\#2 
(left) and MODEL\#3 (right). Lighter colours indicate higher densities of stars. The
simulated diagrams have been built with the 10\%-higher parameters (Table~\ref{tab1}).}
\label{fig3}
\end{figure}

For practical reasons, the age grid is restricted to \ta = 0.2, 1, 3, 5, 8, 10, 20 and 
30\,Myr. Indeed, as can be inferred from Fig.~\ref{fig1}, a finer age grid would be 
redundant. The stellar mass/luminosity relation is taken from the respective merged
set of solar-metallicity Padova and Siess isochrones.
Magnitudes for stars with intermediate age values are obtained by interpolation among the 
neighbouring isochrones. According to the adopted isochrone sets, the minimum stellar mass
is 0.1\,\ms, while the maximum ranges from 60\,\ms\ (at 0.2\,Myr), 36\,\ms\ (5\,Myr), 
19\,\ms\ (10\,Myr), and 9\,\ms\ (30\,Myr).

As described above, the {\em brute-force} nature of our approach tends to be very time
consuming. For instance, a typical simulation for a cluster with $\sim500$ stars would
require a parameter grid composed (at least) of 21 mass bins, 5 ages, 21 distance moduli, 
21 foreground reddenings and 21 differential reddenings. Then, including the $N_{sim}$
clusters in the simulation, the runtime (for the minimum grid) is about 10 hours on a 
single core of an {\em Intel Core i7 920@2.67\,GHz} processor.

\section{Control experiments}
\label{ConExp}

Before turning to actual cases, we apply the approach described above to template 
CMDs built with typical parameters found in PMS-rich young clusters. The main reason 
is to examine the ability to recover the input parameters and their relation with the 
\x2\ statistics. The relevant parameters of the adopted models cover the ranges 
$100\le\Mcl(\ms)\le500$, $3\le\ta(\rm Myr)\le20$, $9.5\le\mMJ\le12$, $0.1\le\eJK\le0.4$,
and $0.5\le\dV\le3.5$. The models are described in Table~\ref{tab1}, which also contains 
quantitative details of the search for solutions on the \x2\ maps. 

Besides the single solution for the absolute \x2\ minimum, we also explore the solutions 
with \x2\ occurring within thresholds 5\%, 10\%, 25\%, and 50\% higher than the absolute 
minimum. For these, the parameters given in Table~\ref{tab1} correspond to the weighted average 
of all the solutions matching each threshold. As weight for each solution we take the individual 
value of $1/R^2_{RMS}$. Except for a few cases, there's little difference among the average 
values of a given model, from the absolute minimum to the 50\% threshold. As expected, the most 
noticeable difference in the output lies in the significantly increasing dispersion around the 
average for higher thresholds. Although somewhat subjective, we believe that the {\em best 
convergence} is obtained 
with the 10\% threshold (which also produces realistic errors), which is consistent with 
the somewhat irregular topology of the \x2\ maps (see below). Overall, the approach seems 
very sensitive to the parameters, especially the age and differential reddening. The 
residual \x2\ of the 10\%-higher solutions occur in the range $0.023\le\x2\le0.051$.
As an additional perspective on the statistical significance of the solutions within the 
adopted residual thresholds, we also provide in Table~\ref{tab1} the percentage of $N_{sol}$
with respect to the full range of possibilities (which depends on the number of parameter
bins). Note that even allowing for the 10\%-higher threshold, the fraction of acceptable 
solutions corresponds to $\sim1 - 5\%$ of the total number. 

\begin{table*}
\caption[]{Recovery of model fundamental parameters ($N_{sim}=500$; $\fB=0.0$)}
\label{tab1}
\renewcommand{\tabcolsep}{2.4mm}
\renewcommand{\arraystretch}{1.25}
\begin{tabular}{rccccccccc}
\hline\hline
$\x2$ &Range&$N_{sol}$&$f_{sol}$&$M_{clu}$& Age &\mMJ&\eJK&\dV&$M_{CMD}$\\
      &     &         &  (\%)   &(\ms)  &(Myr)&(mag)& (mag)& (mag) &(\ms)  \\
(1)&(2)&(3)&(4)&(5)&(6)&(7)&(8)&(9)&(10)\\
\hline
\multicolumn{4}{c}{Input parameters of MODEL\#1}& $250$& $3$&$9.5$ & $0.1$ &$3.0$\\
\hline
$0.0463$   &Abs. minimum  &   1&$1.7\times10^{-3}$&$240\pm0$  &  $3.0\pm0.0 $& $9.60\pm0.00 $& $0.14\pm0.00 $& $2.8\pm0.0 $& $160\pm0$ \\
$\le0.0487$&5\% higher    & 303&$5.1\times10^{-1}$&$233\pm11$ &  $3.0\pm0.1 $& $9.59\pm0.04 $& $0.14\pm0.02 $& $2.8\pm0.2 $& $156\pm8$\\
$\le0.0510$&10\%  higher  &1030&$1.7$             &$235\pm16$ &  $3.0\pm0.1 $& $9.58\pm0.05 $& $0.13\pm0.02 $& $2.9\pm0.2 $& $156\pm10$\\
$\le0.0579$&25\%  higher  &4002&$6.7$             &$234\pm20$ &  $3.0\pm0.1 $& $9.57\pm0.08 $& $0.13\pm0.03 $& $2.9\pm0.2 $& $155\pm13$\\
$\le0.0695$&50\%  higher  &9023&$15.0$            &$234\pm21$ &  $3.0\pm0.2 $& $9.56\pm0.11 $& $0.12\pm0.03 $& $2.9\pm0.2 $& $155\pm14$\\
\hline
\multicolumn{4}{c}{Input parameters of MODEL\#2:}& $100$& $5$&$10.0$ & $0.2$ &$0.5$\\
\hline
$0.0336$   &Abs. minimum &    1&$1.4\times10^{-3}$&$100\pm0$  &  $5.0\pm0.0 $& $10.10\pm0.00 $& $0.22\pm0.00 $& $0.2\pm0.0 $& $106\pm0$ \\
$\le0.0353$&5\% higher   &  214&$3.0\times10^{-1}$&$104\pm6$  &  $5.0\pm0.1 $& $10.08\pm0.04 $& $0.21\pm0.02 $& $0.4\pm0.2 $& $103\pm6$\\
$\le0.0369$&10\%  higher &  828&$1.2$             &$104\pm9$  &  $5.0\pm0.1 $& $10.07\pm0.06 $& $0.20\pm0.02 $& $0.5\pm0.3 $& $101\pm9$\\
$\le0.0420$&25\%  higher & 8977&$13.0$            &$108\pm14$ &  $6.3\pm1.5 $& $ 9.95\pm0.17 $& $0.20\pm0.03 $& $0.5\pm0.3 $& $106\pm14$\\
$\le0.0504$&50\%  higher &28064&$40.0$            &$106\pm15$ &  $6.5\pm1.5 $& $ 9.95\pm0.19 $& $0.21\pm0.03 $& $0.5\pm0.3 $& $103\pm15$\\
\hline
\multicolumn{4}{c}{Input parameters of MODEL\#3:}& $500$& $10$&$12.0$ & $0.3$ &$3.5$\\
\hline
$0.0279$   &Abs. minimum &    1&$5.1\times10^{-4}$&$495\pm0$  &  $8.0\pm0.0 $& $12.20\pm0.00 $& $0.30\pm0.00 $& $3.7\pm0.0 $& $261\pm0$ \\
$\le0.0293$&5\% higher   & 1811&$9.2\times10^{-1}$&$494\pm22$ &  $8.4\pm0.8 $& $12.16\pm0.09 $& $0.31\pm0.02 $& $3.6\pm0.2 $& $265\pm11$\\
$\le0.0307$&10\%  higher & 5871&$3.0$             &$492\pm23$ &  $8.8\pm1.0 $& $12.12\pm0.13 $& $0.31\pm0.02 $& $3.6\pm0.2 $& $265\pm12$\\
$\le0.0349$&25\%  higher &21327&$11.0$            &$490\pm24$ &  $8.8\pm1.2 $& $12.11\pm0.17 $& $0.31\pm0.03 $& $3.6\pm0.2 $& $263\pm13$\\
$\le0.0418$&50\%  higher &51482&$26.0$            &$489\pm24$ &  $8.8\pm1.5 $& $12.12\pm0.20 $& $0.30\pm0.04 $& $3.6\pm0.2 $& $263\pm13$\\
\hline
\multicolumn{4}{c}{Input parameters of MODEL\#4:}& $150$& $20$&$11.0$ & $0.4$ &$1.5$\\
\hline
$0.0209$   &Abs. minimum &    1&$6.1\times10^{-4}$&$120\pm0$  &  $20.0\pm0.0 $& $11.10\pm0.00 $& $0.46\pm0.00 $& $1.1\pm0.0 $& $94\pm0$ \\
$\le0.0219$&5\% higher   & 2154&$1.3$             &$142\pm19$ &  $24.3\pm4.9 $& $11.09\pm0.06 $& $0.44\pm0.03 $& $1.5\pm0.3 $& $102\pm14$\\
$\le0.0230$&10\%  higher & 8735&$5.3$             &$142\pm20$ &  $24.5\pm5.0 $& $11.09\pm0.10 $& $0.44\pm0.03 $& $1.5\pm0.3 $& $101\pm14$\\
$\le0.0261$&25\%  higher &43299&$26.0$            &$138\pm23$ &  $24.6\pm5.5 $& $11.07\pm0.19 $& $0.43\pm0.04 $& $1.5\pm0.3 $& $99\pm16$\\
$\le0.0313$&50\%  higher &89775&$55.0$            &$135\pm24$ &  $24.3\pm6.1 $& $11.07\pm0.22 $& $0.42\pm0.05 $& $1.5\pm0.3 $& $97\pm17$\\
\hline
\end{tabular}
\begin{list}{Table Notes.}
\item Cols.~(1) and (2): \x2\ value and corresponding range with respect to the absolute 
minimum; Col.~(3): number of solutions occurring within the \x2\ range; Col.~(4): percentage
of $N_{sol}$ with respect to the full range of solutions; Col.~(5): actual cluster 
mass; Col.~(9): differential reddening; Col.~(10): mass 
detected in the CMD. $N_{sim}$ is the number of simulated clusters of mass \Mcl\ and age
\ta. The average stellar mass of the models is $\overline{m_*}\approx0.6\,\ms$.
\end{list}
\end{table*}

Selected two-dimensional projections of $\x2=\x2(\Mcl, \ta, \dV, \eJK, \mMJ)$ are shown
in Fig.~\ref{fig2} for MODELS\#2 and 3, those that present extreme values of cluster mass 
and differential reddening. As discussed above, the approach presents conspicuous convergence 
towards the input values of the age, cluster mass, and apparent distance modulus. However, 
the \x2\ maps present some spread in both foreground and differential reddening values. In 
particular, there clearly is an anti-correlation between both reddening sources, in the sense 
that low (or high) foreground reddenings are compensated for by the approach with high (or 
low) values of differential reddening.

An important information that in principle could be obtained from CMDs is the cluster mass.
More specifically, the fraction stored in MS and PMS stars (in this work it means the mass 
sum for all stars more massive than $\ge0.1\,\ms$). However, as discussed in Sect.~\ref{Intro}, 
the intrinsic age spread together with the presence of differential reddening, complicate 
the task of finding the mass of each star in a CMD. Besides, because of limitations inherent 
to any photometric system (which, in the present work, are simulated by means of the quality 
criterion (step {\em (viii)} in Sect.~\ref{Test}), the number of stars that remain detectable 
in a CMD tends to decrease as the distance modulus increases. Consequently, the same applies 
to the stellar mass that would be measured in a CMD ($M_{CMD}$) with respect to the actual cluster 
mass (\Mcl). After derivation of the fundamental parameters, $M_{CMD}$ can be estimated by finding
the probable mass for each star in the CMD. This can be done by interpolation (of the observed 
colour and magnitude of each star) among the nearest isochrones (for instance, Fig.~\ref{fig6}), 
but the final result would depend heavily on the amount of differential reddening, photometric 
noise, etc. Obviously, the presence of a large fraction of binaries would produce low values of 
$M_{CMD}$. By construction, the present approach provides directly the actual cluster mass and, 
since we keep track of the mass 
(and photometry) of each star that is used in the simulations, we can also compute the mass 
present in a CMD. Both mass values are given in Table~\ref{tab1}. Interestingly, for the 
range of \mMJ\ covered by the models, the actual cluster mass and that present in the CMDs 
are essentially the same. 

\begin{figure}
\resizebox{\hsize}{!}{\includegraphics{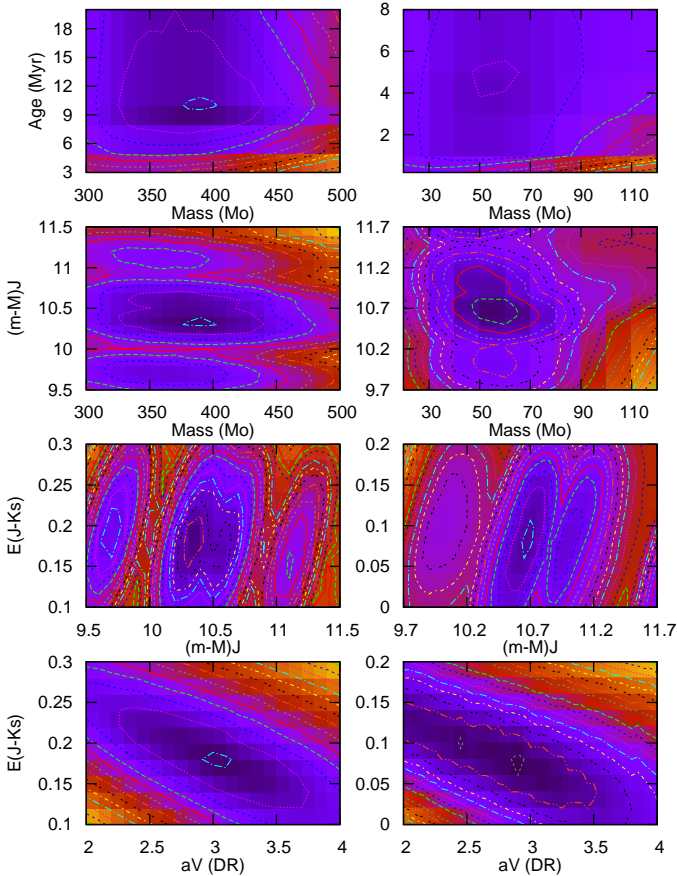}}
\caption[]{Same as Fig.~\ref{fig2} for the young clusters Collinder\,197 (left)
and Pismis\,5 (right). The absolute minima are $\x2=0.054$ (Collinder\,197) and 
$\x2=0.019$ (Pismis\,5).}
\label{fig4}
\end{figure}

\begin{figure}
\resizebox{\hsize}{!}{\includegraphics{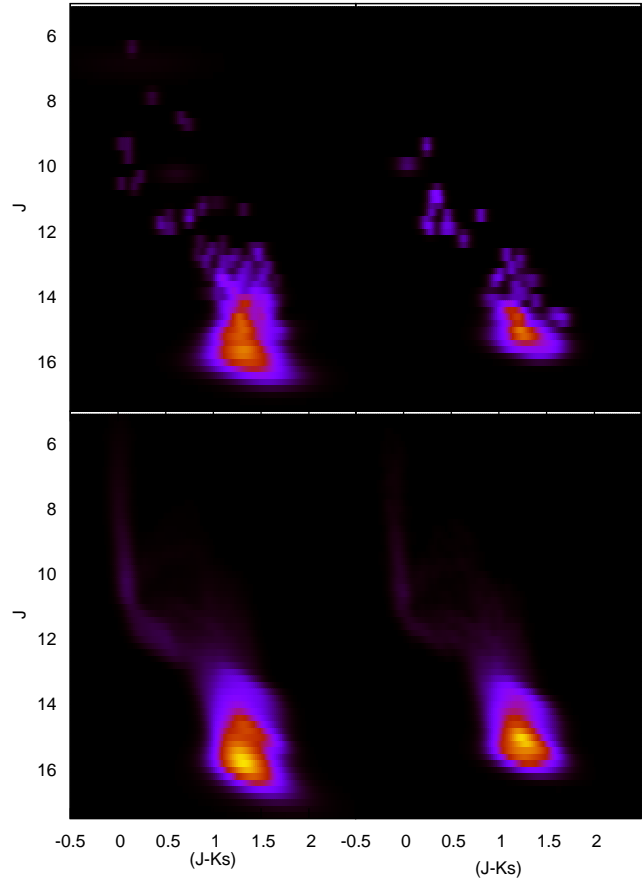}}
\caption[]{Same as Fig.~\ref{fig3} for the field-star decontaminated (top panels) and 
simulated (bottom) Hess diagrams of Collinder\,197 (left) and Pismis\,5 (right). }
\label{fig5}
\end{figure}

Another quality assessment criterion is provided by the compared morphology between
the model and simulated Hess diagrams (Fig.~\ref{fig3}). The latter have been
constructed with the parameters found with the 10\% threshold (Table~\ref{tab1}). 
Given the low-mass nature of the models considered here, some discreteness in their 
Hess diagrams is expected. On the other hand, the simulated Hess diagrams correspond 
to the average over $N_{sim}$ simulated clusters (of mass \Mcl\ and age \ta), and 
thus, they present a smoother distribution. Nevertheless, model and simulated 
Hess diagrams present a good correspondence.

\begin{table*}
\caption[]{Fundamental parameters of the young clusters Collinder\,197 and Pismis\,5}
\label{tab2}
\renewcommand{\tabcolsep}{1.85mm}
\renewcommand{\arraystretch}{1.25}
\begin{tabular}{rcccccccccc}
\hline\hline
$\x2$ &Range&$N_{sol}$&$f_{sol}$&$M_{clu}$& Age &\mMJ&\eJK&\dV&$M_{CMD}$&\ds\\
      &     &         &  (\%)   &(\ms)  &(Myr)&(mag)&(mag)& (mag) &(\ms)& (kpc)  \\
(1)&(2)&(3)&(4)&(5)&(6)&(7)&(8)&(9)&(10)&(11)\\
\hline
   &\multicolumn{10}{c}{Collinder\,197 ($\fB = 0.0$) }\\
\hline
$0.0546$  &Abs. min.   &    1&$3.6\times10^{-4}$& $390\pm0$  &  $10.0\pm0.0 $& $10.30\pm0.00 $& $0.18\pm0.00 $& $3.1\pm0.0 $& $245\pm0$ &$1.0\pm0.0$ \\
$\le0.0574$&5\%  higher&  186&$6.7\times10^{-2}$& $392\pm18$ &  $ 9.8\pm0.6 $& $10.37\pm0.09 $& $0.18\pm0.02 $& $3.0\pm0.2 $& $249\pm11$&$1.0\pm0.1$\\
$\le0.0601$&10\% higher&  759&$2.7\times10^{-1}$& $393\pm24$ &  $ 9.5\pm0.9 $& $10.41\pm0.12 $& $0.18\pm0.03 $& $3.0\pm0.3 $& $248\pm15$&$1.0\pm0.1$\\
$\le0.0683$&25\% higher& 6527&$2.4$             & $388\pm36$ &  $ 9.8\pm2.7 $& $10.42\pm0.22 $& $0.18\pm0.04 $& $3.0\pm0.5 $& $245\pm23$&$1.0\pm0.1$\\
$\le0.0819$&50\% higher&28754&$10.0$            & $381\pm41$ &  $ 9.9\pm4.0 $& $10.48\pm0.35 $& $0.18\pm0.05 $& $3.0\pm0.5 $& $241\pm26$&$1.1\pm0.2$\\
\hline
   &\multicolumn{10}{c}{Collinder\,197 ($\fB = 0.5$) }\\
\hline
$0.0566$   &Abs. min.  &    1&$2.0\times10^{-4}$& $660\pm0$  &  $10.0\pm0.0 $& $10.80\pm0.00 $& $0.18\pm0.00 $& $3.1\pm0.0 $& $351\pm0$ &$1.2\pm0.0$ \\
$\le0.0595$&5\%  higher&  354&$6.9\times10^{-2}$& $666\pm30$ &  $10.0\pm0.2 $& $10.85\pm0.07 $& $0.19\pm0.02 $& $3.1\pm0.2 $& $354\pm16$&$1.3\pm0.1$\\
$\le0.0623$&10\% higher& 2037&$4.0\times10^{-1}$& $661\pm39$ &  $ 9.9\pm2.3 $& $10.88\pm0.21 $& $0.19\pm0.03 $& $3.1\pm0.3 $& $354\pm21$&$1.3\pm0.1$\\
$\le0.0708$&25\% higher&22660&$4.4$             & $648\pm53$ &  $10.9\pm4.7 $& $10.86\pm0.42 $& $0.19\pm0.04 $& $3.1\pm0.5 $& $347\pm28$&$1.3\pm0.2$\\
$\le0.0849$&50\% higher&84596&$17.0$            & $645\pm57$ &  $10.6\pm5.2 $& $10.96\pm0.49 $& $0.19\pm0.05 $& $3.0\pm0.6 $& $345\pm31$&$1.3\pm0.3$\\
\hline
   &\multicolumn{10}{c}{Collinder\,197 ($\fB = 1.0$) }\\
\hline
$0.0553$  &Abs. min.   &    1&$2.0\times10^{-4}$& $820\pm0$  &  $10.0\pm0.0 $& $11.20\pm0.00 $& $0.20\pm0.00 $& $3.0\pm0.0 $& $405\pm0$ &$1.5\pm0.0$\\
$\le0.0580$&5\%  higher&  359&$7.0\times10^{-2}$& $824\pm35$ &  $ 9.6\pm0.8 $& $11.21\pm0.11 $& $0.19\pm0.02 $& $3.1\pm0.2 $& $404\pm17$&$1.5\pm0.1$\\
$\le0.0608$&10\% higher& 1697&$3.3\times10^{-1}$& $816\pm45$ &  $ 9.3\pm1.0 $& $11.27\pm0.14 $& $0.19\pm0.02 $& $3.1\pm0.3 $& $402\pm22$&$1.5\pm0.1$\\
$\le0.0691$&25\% higher&12174&$2.4$             & $803\pm55$ &  $ 9.9\pm3.1 $& $11.25\pm0.27 $& $0.19\pm0.04 $& $3.1\pm0.5 $& $395\pm27$&$1.5\pm0.2$\\
$\le0.0829$&50\% higher&54465&$11.0$            & $796\pm58$ &  $10.5\pm4.8 $& $11.27\pm0.45 $& $0.20\pm0.05 $& $3.1\pm0.5 $& $390\pm28$&$1.5\pm0.3$\\
\hline
\hline
   &\multicolumn{10}{c}{Pismis\,5 ($\fB = 0.0$)}\\
\hline
$0.0188$   &Abs. min.&       1&$3.5\times10^{-4}$& $60\pm0$  &  $5.0\pm0.0 $& $10.75\pm0.00 $& $0.08\pm0.00 $& $2.9\pm0.0 $& $37\pm0$ &$1.3\pm0.0$ \\
$\le0.0197$&5\%  higher  &1301&$4.5\times10^{-1}$& $61\pm6$  &  $5.0\pm0.1 $& $10.75\pm0.08 $& $0.09\pm0.03 $& $2.9\pm0.3 $& $38\pm4$ &$1.3\pm0.1$\\
$\le0.0207$&10\% higher & 4816&$1.7$             & $62\pm10$ &  $5.0\pm0.9 $& $10.79\pm0.14 $& $0.09\pm0.03 $& $2.9\pm0.3 $& $39\pm6$ &$1.3\pm0.1$\\
$\le0.0235$&25\% higher& 51006&$18.0$            & $68\pm18$ &  $5.6\pm1.9 $& $10.88\pm0.26 $& $0.11\pm0.04 $& $2.9\pm0.3 $& $42\pm11$&$1.4\pm0.2$\\
$\le0.0282$&50\% higher&133571&$46.0$            & $67\pm21$ &  $5.8\pm1.9 $& $10.90\pm0.31 $& $0.11\pm0.04 $& $2.9\pm0.3 $& $41\pm13$&$1.4\pm0.2$\\
\hline
   &\multicolumn{10}{c}{Pismis\,5 ($\fB = 0.5$)}\\
\hline
$0.0176$   &Abs. min.&       1&$3.8\times10^{-4}$& $110\pm0$  &  $4.0\pm0.0 $& $11.75\pm0.00 $& $0.10\pm0.00 $& $3.0\pm0.0 $& $57\pm0$ &$2.1\pm0.0$ \\
$\le0.0185$&5\%  higher  &1395&$5.3\times10^{-1}$& $111\pm8$  &  $4.0\pm0.2 $& $11.80\pm0.09 $& $0.10\pm0.02 $& $3.0\pm0.3 $& $58\pm4$ &$2.1\pm0.1$\\
$\le0.0194$&10\% higher & 6156&$2.3$             & $110\pm13$ &  $3.7\pm0.7 $& $11.89\pm0.23 $& $0.10\pm0.03 $& $3.1\pm0.3 $& $56\pm6$ &$2.2\pm0.2$\\
$\le0.0220$&25\% higher& 52449&$20.0$            & $119\pm22$ &  $4.6\pm1.4 $& $11.82\pm0.29 $& $0.11\pm0.04 $& $3.1\pm0.3 $& $62\pm11$&$2.1\pm0.3$\\
$\le0.0264$&50\% higher&140800&$54.0$            & $121\pm23$ &  $5.3\pm2.0 $& $11.82\pm0.31 $& $0.11\pm0.04 $& $3.1\pm0.3 $& $58\pm11$&$2.1\pm0.3$\\
\hline
   &\multicolumn{10}{c}{Pismis\,5 ($\fB = 1.0$)}\\
\hline
$0.0189$   &Abs. min.&       1&$3.2\times10^{-4}$& $140\pm0$  &  $5.0\pm0.0 $& $11.60\pm0.00 $& $0.10\pm0.00 $& $3.1\pm0.0 $& $67\pm0$ &$1.9\pm0.0$ \\
$\le0.0198$&5\%  higher  &4865&$1.6$             & $131\pm14$ &  $4.3\pm1.0 $& $11.74\pm0.20 $& $0.10\pm0.02 $& $3.1\pm0.3 $& $62\pm7$ &$2.0\pm0.2$\\
$\le0.0208$&10\% higher &20900&$6.7$             & $131\pm18$ &  $4.1\pm1.0 $& $11.81\pm0.26 $& $0.10\pm0.03 $& $3.2\pm0.3 $& $62\pm9$ &$2.1\pm0.3$\\
$\le0.0236$&25\% higher& 95284&$30.0$            & $134\pm23$ &  $4.6\pm1.7 $& $11.84\pm0.32 $& $0.11\pm0.04 $& $3.2\pm0.3 $& $63\pm11$&$2.1\pm0.3$\\
$\le0.0283$&50\% higher&183790&$59.0$            & $132\pm24$ &  $4.9\pm2.0 $& $11.88\pm0.34 $& $0.11\pm0.04 $& $3.2\pm0.3 $& $61\pm11$&$2.2\pm0.3$\\
\hline
\end{tabular}
\begin{list}{Table Notes.}
\item Collinder\,197 presents 690 stars in the CMD, while Pismis\,5 shows only 101. 
Collinder\,197 was analysed with $N_{sim}=250$ simulated clusters and, for equivalent 
statistical results, Pismis\,5 was analysed with $N_{sim}=1000$. \fB\ is the binary
fraction. The average stellar mass of the model clusters is $\overline{m_*}\approx0.6\,\ms$.
\end{list}
\end{table*}

\section{Application to actual young clusters}
\label{TCADF}

Having demonstrated the convergence efficiency of our approach with artificial clusters
(Sect.~\ref{ConExp}), we now move on to examining properties of actual cases. For this 
we have selected two young clusters previously studied by our group, Collinder\,197 
(\citealt{vdB92}) and Pismis\,5 (\citealt{Pi5}). Both present typical CMDs of young
clusters, with the difference that Collinder\,197 has about 10 times more stars 
(essentially PMS) in the CMD than Pismis\,5. 

To put the results derived in the present paper in context, we provide here a brief
explanation of the previous method used by our group to obtain parameters of both 
clusters. While the procedure used in \citet{vdB92} and \citet{Pi5} to construct the 
field-star decontaminated CMDs follows a quantitative approach, the fundamental 
parameter derivation used in both cases is somewhat subjective, depending essentially 
on a qualitative assessment of the differential reddening. Specifically, the MS$+$PMS 
isochrones (for a range of ages) are set to zero distance modulus and foreground 
reddening, and shifted in magnitude and colour until they produce a satisfactory fit 
of the blue border (and redwards spread) of the MS and PMS stellar distribution. 
Since we did not dispose of any quantitative tool to evaluate the differential 
reddening, we simply assumed it to be equivalent to the size of the reddening vector
that matched the colour spread among the lower-PMS sequences. For the (CMD extracted 
from the) region within $R=10\arcmin$ from the 
centre of Collinder\,197, \citet{vdB92} find $\ta=5\pm4$\,Myr, $\mMJ=10.4\pm0.4$, 
$\eJK=0.17\pm0.08$, $M_{CMD}\approx450\pm100\,\ms$, and the distance from the Sun 
$\ds=1.1\pm0.2$\,kpc. For $R=6\arcmin$ of Pismis\,5, \citet{Pi5} find $\ta=5\pm4$\,Myr, 
$\mMJ=10.4\pm0.1$, $\eJK=0.20\pm0.02$, and $M_{CMD}\approx58\pm8\,\ms$, and 
$\ds=1.0\pm0.1$\,kpc. The adopted age (and uncertainty) is simply the average (and 
half the difference) between the youngest and oldest MS$+$PMS isochrones compatible 
with the CMD morphology. The mass was estimated by multiplying the number of CMD 
stars by the average stellar mass (based on \citealt{Kroupa2001} MF); thus, it 
represents the CMD mass.

Considering the above features, both clusters are excellent candidates to apply the 
present approach. Given the low number of stars (101) in the CMD of Pismis\,5, we 
increased the number of simulated clusters to $N_{sim}=1000$ for statistically more 
significant results. Collinder\,197 has 690 stars in the CMD, so, we kept $N_{sim}=250$. 
For a more comprehensive analysis, we have applied our approach considering the 
binary fractions $\fB=0,~0.5$ and $1$. The fundamental parameters obtained with the 
present approach are given in Table~\ref{tab2}.

\begin{figure}
\resizebox{\hsize}{!}{\includegraphics{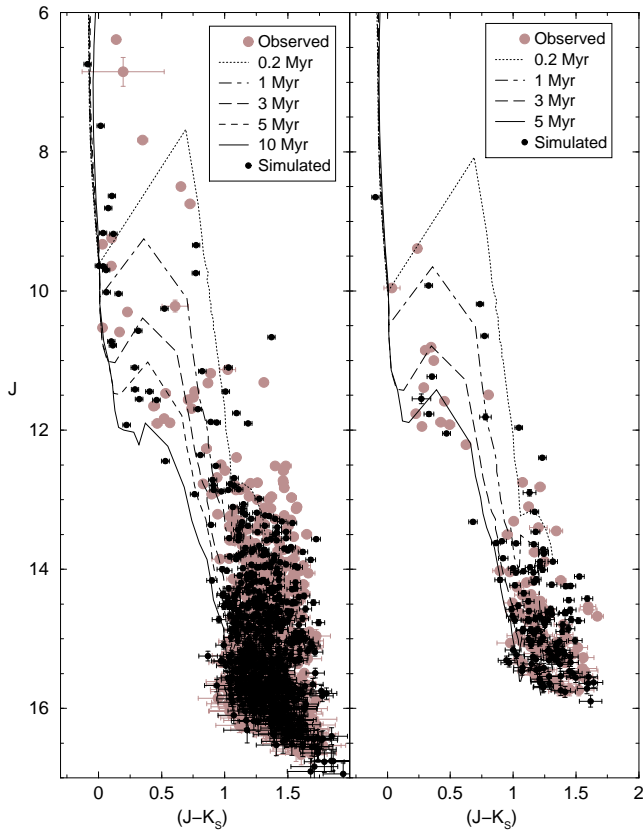}}
\caption[]{Observed CMDs of Cr\,197 (left) and Pismis\,5 (right) compared to a single
realisation taken from the respective simulations. The extraction radii are 10\arcmin\
(Cr\,197) and 6\arcmin\ (Pismis\,5).}
\label{fig6}
\end{figure}

Most of the parameters are essentially insensitive to the presence of unresolved
binaries in the CMD. This is especially true for the age, foreground reddening and, to a lesser 
degree, differential reddening. On the other hand, the most remarkable difference between the 
presence of (CMD unresolved) binary fractions of 0\% and 100\% lies in the cluster mass, with 
$\fB=1$ requiring $\approx$ twice more mass than for $\fB=0$. Next, the binary-related brightening 
of the stellar sequences (for $\fB=1$) requires an apparent distance modulus, on average, 0.84 and 
0.96 higher than for $\fB=0$, respectively for Collinder\,197 and Pismis\,5. This reflects on 
distances to the Sun $\approx50\%$ higher if $\fB=1$. Within the uncertainties, the parameters 
corresponding to $\fB=0.5$ tend to be closer to those obtained with $\fB=1$. Also, for a given
binary fraction, the parameters are essentially unchanged with respect to the \x2\ ranges 
considered, although with an increasing dispersion around the average for high \x2. Finally, the 
convergence level of the approach, as measured by \x2, is essentially the same for the binary 
fractions considered here. In this sense, our approach appears to be insensitive to the binary 
fraction, at least for clusters with a significant age spread.

Irrespective of the binary fraction, our results confirm that Pismis\,5 is indeed a very-low 
mass ($\Mcl\sim60 - 140\,\ms$) and young ($\ta\sim4 - 6$\,Myr) cluster, affected by a moderate 
amount of differential reddening ($\dV\sim3.0$), and a low foreground reddening ($\aV\sim0.6$).
Such a low \aV, together with the moderate apparent distance modulus ($\mMJ\sim10.8$, for $\fB=0$), 
puts Pismis\,5 at a distance from the Sun of $\ds=1.3\pm0.1$\,kpc. However, this distance may
be significantly higher ($\ds=2.1\pm0.3$\,kpc) if $\fB=1$.

As anticipated, Collinder\,197 is more massive ($\Mcl\sim390-820\,\ms$) and somewhat older 
($\ta\sim9.5$\,Myr) than Pismis\,5, also with a similar value of differential reddening
($\dV\sim3.0$), and a somewhat higher foreground reddening ($\aV\sim1.2$). The distance 
moduli $\mMJ\sim10.4$ ($\fB=0$) and $\mMJ\sim11.3$ ($\fB=1$) imply the distances 
$\ds=1.0\pm0.1$\,kpc and $\ds=1.5\pm0.1$\,kpc, respectively. Under the assumptions adopted
in Sect.~\ref{Test}, the star-formation age spread of Collinder\,197 is about twice that of 
Pismis\,5. Interestingly, despite the subjectiveness of the (previous) qualitative assessment 
we applied to Collinder\,197 and Pismis\,5, the results of both methods are somewhat compatible, 
within the uncertainties.

The \x2\ projections of Collinder\,197 and Pismis\,5 (Fig.~\ref{fig4}) present similar 
convergence patterns as those of the control tests (Fig.~\ref{fig2}), including the 
foreground and differential reddening anti-correlation. The residual \x2\ of the 10\%-higher 
solutions of Pismis\,5 ($\x2\le0.021$) is even lower than those of the control tests, and 
somewhat higher ($\x2\le0.06$) for Collinder\,197. However, we point out that part of this may
be linked to the fact that Collinder\,197 has $\approx7$ times more stars in the CMD, which
can result in a larger sum of the residuals. Next, considering the low number of free
parameters used by our approach, the observed and simulated Hess diagrams (Fig.~\ref{fig5}) 
present a satisfactory correspondence. Finally, in Fig.~\ref{fig6} we compare a single CMD 
realisation - randomly selected among the $N_{sim}$ simulated clusters - with the observed 
CMDs. Having in mind the fact that both are low-mass, young clusters, some differences are 
more evident because of the low number of stars, especially in the MS, when an isolated, 
random CMD is used to illustrate the simulations. This effect should be particularly noticeable 
in poorly-populated clusters, such as Pismis\,5. Nevertheless, simulated and observed CMDs 
are similar, in both cases.

\section{Summary and conclusions}
\label{Conclu}

In this paper we describe an approach based on CMDs and near-infrared photometry to 
obtain more accurate fundamental parameters of young star clusters. However, since the 
presence of large fractions of circumstellar disks around PMS stars can lead to significant 
excesses in the near-infrared (e.g. \citealt{Meyer97}), the applicability of our approach 
may be restricted to clusters older than $\sim3$\,Myr (e.g. \citealt{HLL01}). Given its
statistical nature, our method should be very efficient for well populated clusters,
which give rise to smoother stellar-density distributions in the Hess diagrams. But more
interestingly, it is expected to work also with the low-mass clusters with CMDs dominated 
by PMS stars and significantly affected by differential reddening. Among these parameters, 
cluster mass and age are important to understand the dynamical state (e.g. \citealt{DetAnal}) 
of clusters, especially those undergoing the rapidly changing and potentially-destructive 
evolutionary phase characterised by the first few $10^7$\,yr.

The approach involves simulating the effect of (random) differential reddening on
MS and PMS stars, and its observable results on CMDs. In this paper we apply it to 
CMDs of star clusters simulated in the 2MASS photometric system. Obviously, it
can be adapted to any photometric system, provided isochrones for low-mass stars and
very young ages are available.

Beginning with a given cluster mass (\Mcl) and age (\ta), we distribute the individual stellar 
masses and magnitudes according to the \citet{Kroupa2001} MF and an MS$+$PMS isochrone set. 
Subsequently, we build the corresponding simulated Hess diagram for different values of the 
apparent distance modulus (\mMJ), differential (\dV) and foreground (\eJK) reddening values. 
We repeat this for a range of cluster mass and age, searching for the best match between the 
simulated and observed Hess diagrams. The best-match parameters are searched around the minima
of the hypersurface defined by the root mean squared residual $\x2=\x2(\Mcl, \ta, \dV, \eJK, 
\mMJ)$.

Tests with model clusters containing parameters typical of objects undergoing such 
an early phase have shown that the present tool converges to the input values, especially 
when one allows for solutions that occur with \x2\ 10\% higher than the absolute minimum.
We also investigate how unresolved binaries affect the derived parameters. Compared 
to a CMD containing only single stars, even assuming a presence of 100\% of unresolved 
binaries has little effect (to within the uncertainties) on cluster age, foreground and 
differential reddenings. Significant differences occur in the cluster mass  and distance 
from the Sun. About twice more mass (in individual stars) is required for $\fB=1$ than 
for $\fB=0$, and because of the relative brightening due to the binaries, the distance
from the Sun is $\approx50\%$ larger for $\fB=1$ than for $\fB=0$.

As a caveat we note that we consider here a particular isochrone set. Thus, the results 
of our approach tend to be model dependent, since different isochrone sets may lead to 
different values of mass and age of individual stars (e.g. \citealt{Hill08}) and, 
consequently, star clusters. In addition, we minimise the number of free parameters by assuming
uniform (or flat) distributions of differential reddening and stellar age. Both conditions
may be partly unrealistic, especially the latter for clusters older than $\sim10$\,Myr, 
since it would imply a slow and steady star formation rate. However, the clusters we use
here to test the approach, Collinder\,197 and Pismis\,5, are younger than $\sim10$\,Myr.
The relative low-mass nature of both clusters may imply CMDs lacking constraints to find 
the {\em best values} for a large number of free parameters.




The general conclusion is that the inclusion in the simulations of several effects that 
affect the CMD morphology - especially the differential reddening - produces constrained 
values of the fundamental parameters. Thus, when photometry is the only available information,
our approach minimises the subjectiveness associated with the parameter derivation of young 
clusters.

\section*{Acknowledgements}
We thank an anonymous referee for interesting comments and suggestions.
We acknowledge support from the Brazilian Institution CNPq.

\label{lastpage}
\end{document}